%% file: main.tex
\documentclass[article,compsoc]{IEEEtran}

\input{customize}

\begin{document}

\title{ALIGN: A System for Automating Analog Layout
\thanks{This work was supported by the DARPA IDEA program under SPAWAR contract N660011824048.}}

\author{
Tonmoy~Dhar$^1$,
Kishor~Kunal$^1$,
Yaguang~Li$^2$, 
Meghna~Madhusudan$^1$,
Jitesh~Poojary$^1$,  \\
Arvind~K.~Sharma$^1$,
Wenbin~Xu$^2$,
Steven~M.~Burns$^3$,
Ramesh~Harjani$^1$,
Jiang~Hu$^2$, \\
Desmond A. Kirkpatrick$^3$,
Parijat~Mukherjee$^3$,
Sachin~S.~Sapatnekar$^1$, and
Soner~Yaldiz$^3$\\
$^1$ University of Minnesota, Minneapolis, MN, USA \\
$^2$ Texas A\&M University, College Station, TX, USA \\
$^3$ Intel Labs, Hillsboro, OR, USA
}

\maketitle

\begin{abstract}
ALIGN (``Analog Layout, Intelligently Generated from Netlists'') is an
open-source automatic layout generation flow for analog circuits. ALIGN
translates an input SPICE netlist to an output GDSII layout, specific to a
given technology, as specified by a set of design rules.  The flow first
automatically detects hierarchies in the circuit netlist and translates layout
synthesis to a problem of hierarchical block assembly.  At the lowest level,
parameterized cells are generated using an abstraction of the design rules;
these blocks are then assembled under geometric and electrical constraints to
build the circuit layout.  ALIGN has been applied to generate layouts for a
diverse set of analog circuit families: low frequency analog blocks, wireline
circuits, wireless circuits, and power delivery circuits.
\end{abstract}	

\input{intro}
\input{overview}
\input{opensource}

\input{software}

\input{results}

\input{conclusion}

\bibliographystyle{misc/ieeetr2}
\bibliography{bib/main}

\end{document}

%% file: customize.tex
\usepackage{amsmath}
\usepackage{blindtext}
\usepackage{graphicx}
\usepackage{algorithm}
\usepackage{algcompatible}
\usepackage{color}
\usepackage{cite}
\usepackage{url}
\usepackage{subcaption}
\usepackage[justification=centering,font=small,labelfont=bf]{caption}
\definecolor{gray1}{gray}{0.90}
\definecolor{gray2}{gray}{0.98}
\definecolor{light-gray}{gray}{0.95}

\newcommand{\ignore}[1]{}
\newcommand{\redHL}[1]{\textcolor{red}{#1}}
\newcommand{\blueHL}[1]{{\textcolor{black}{#1}}}

\usepackage{tikz}
\usetikzlibrary{shapes,arrows}

%% file: intro.tex
\section{Motivation and Goals}
\label{sec:intro}

ALIGN (Analog Layout, Intelligently Generated from Netlists) is an open-source
layout generator for analog circuits that is currently under development. The
release date for Version 1 of the software flow is in August 2020.  The ALIGN
project engages a joint academic/industry team to translate a SPICE-level
netlist into a physical layout, with 24-hour turnaround and no human in the
loop.  The ALIGN flow inputs a netlist whose topology and transistor sizes have
already been chosen, specifications, and a process design kit (PDK), and
outputs GDSII.  

The philosophy of ALIGN is to compositionally synthesize the layout by
first identifying layout hierarchies in the netlist, then generating
correct-by-construction layouts at the lowest level of hierarchy, and finally
assembling blocks at each level of hierarchy during placement and routing.
Thus, a key step in ALIGN is to identify these hierarchies to recognize the
building blocks of the design. In doing so, ALIGN mimics the human designer,
who identifies known blocks, lays them out, and then builds the overall layout
hierarchically.
At the lowest level of this hierarchy is an individual
transistor; these transistors are then combined into larger fundamental
primitives (e.g., differential pairs, current mirrors), then modules (e.g.,
operational transconductance amplifiers (OTAs)), up through several levels of
hierarchy to the system level (e.g., an RF transceiver).  ALIGN uses a mix of
algorithmic techniques, template-driven design, and machine learning (ML) to
create layouts that are at the level of sophistication of the expert designer. 

\begin{figure}
\centering
\includegraphics[width=0.4\textwidth]{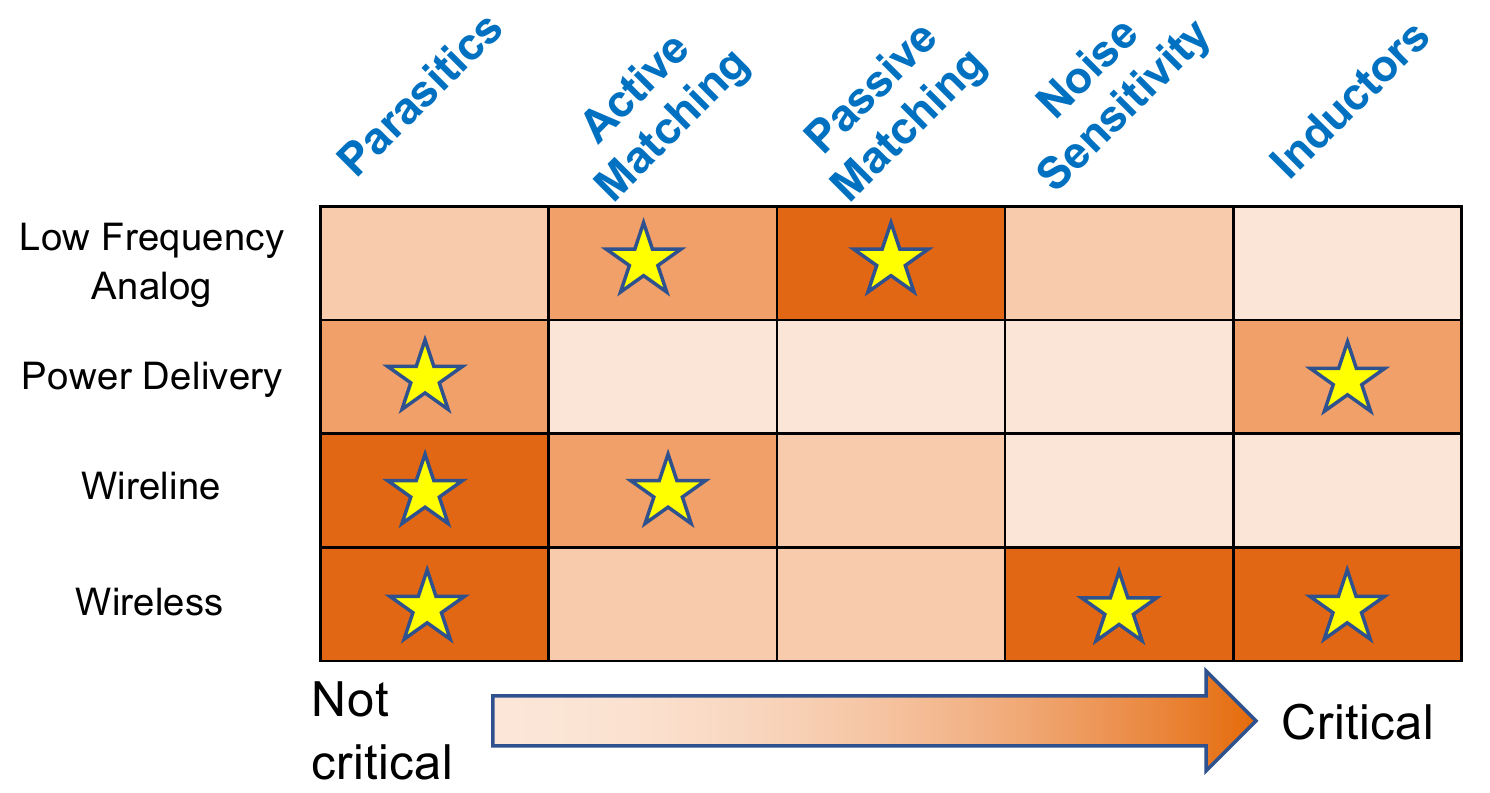}
\caption{Classification of analog circuits, showing the factors that are
important in each category.}
\label{fig:FourCategories}
\end{figure}

Unlike digital designs that are built from a composition of a small number of
building blocks, analog circuits tend to use a wide variety of structures.  Each
of these has its own constraints and requirements, and traditionally only the
expert designer has been able to build circuits that could deliver high performance.
ALIGN targets a wide variety of analog designs, \blueHL{in both bulk and
FinFET technologies, covering four broad classes of functionality}:
\begin{itemize}
\item Low-frequency components that include analog-to-digital converters
(ADCs), amplifiers, and filters.
\item Wireline components that include clock/data recovery, equalizers, and
phase interpolators.
\item RF/Wireless components that implement transmitters, receivers, etc.
\item Power delivery components include capacitor- and inductor-based DC-to-DC
converters.
\end{itemize}
Each class is characterized by similar building blocks that may have a similar
set of performance parameters, although it should be mentioned that there is
considerable diversity even within each class.  An overview of factors that are
important in the design for each category is summarized in
Fig.~\ref{fig:FourCategories}.

There have been several prior efforts to automate analog layout
synthesis~\cite{Harjani89,Cohn91,Graeb10,Eick11,Ou13,Ma11,Wu15}, but these
methods are not widely deployed in tools today. Some methods address limited
classes of designs; others cannot be tuned to handle a wide enough set of
variants of the same design class.  Moreover, there is a general consensus that
prior methods for automating analog layout have been unable to match the expert
designer, both in terms of the ability to comprehend and implement specialized
layout tricks, and the number and variety of topologies with circuit-specific
constraints.  The ultimate goal for analog layout synthesis is to match the
quality of a hand-crafted design.  

In recent years, the landscape has shifted in several ways, making automated
layout solutions attractive.  {\em First}, in nanometer-scale technologies,
restricted design rules with fixed pitches and unidirectional routing limit the
full freedom for layout that was available in older technologies, thus reducing
the design space to be explored during layout, reducing the advantage to the human
expert.  {\em Second}, today more analog blocks are required in integrated
systems than before, and several of these require correct functionality and
modest performance.  The combination of increasing analog content with the
relaxation in specifications creates a sweet spot for analog automation.  Even
for high-performance blocks, an automated layout generator could considerably
reduce the iterations between circuit optimization and layout, where layout
generation is the primary bottleneck.  {\em Third}, the advent of ML provides
the promise for attacking the analog layout problem in a manner that was not
previously possible\blueHL{, and set the stage for no-human-in-the-loop design.}
 
This article provides an overview of the technical details of ALIGN and shows
how ALIGN has been used to translate analog circuit netlists to layouts.
\blueHL{The core ALIGN engine can be run with no human in the loop, enabled by
ML algorithms that perform the functions typically performed by humans, e.g.,
recognizing hierarchies in the circuit during auto-annotation, or generating
symmetry constraints for layout. ML algorithms can also be instrumental in
creating rapid electrical constraint checkers, which verify whether a candidate
placement/routing solution meets performance constraints or not, and using this
to guide the place-and-route engine towards optima that meet all specification.
For deeper details, the reader is referred to detailed descriptions
in~\cite{Kunal20,Kunal20ICCAD,Li20,Li20ICCAD}, and to watch for new
publications of ongoing work by our group.}

%% file: overview.tex
\section{The Technical Core of ALIGN}
\label{sec:overview}

\begin{figure}
\centering
\includegraphics[width=0.5\textwidth]{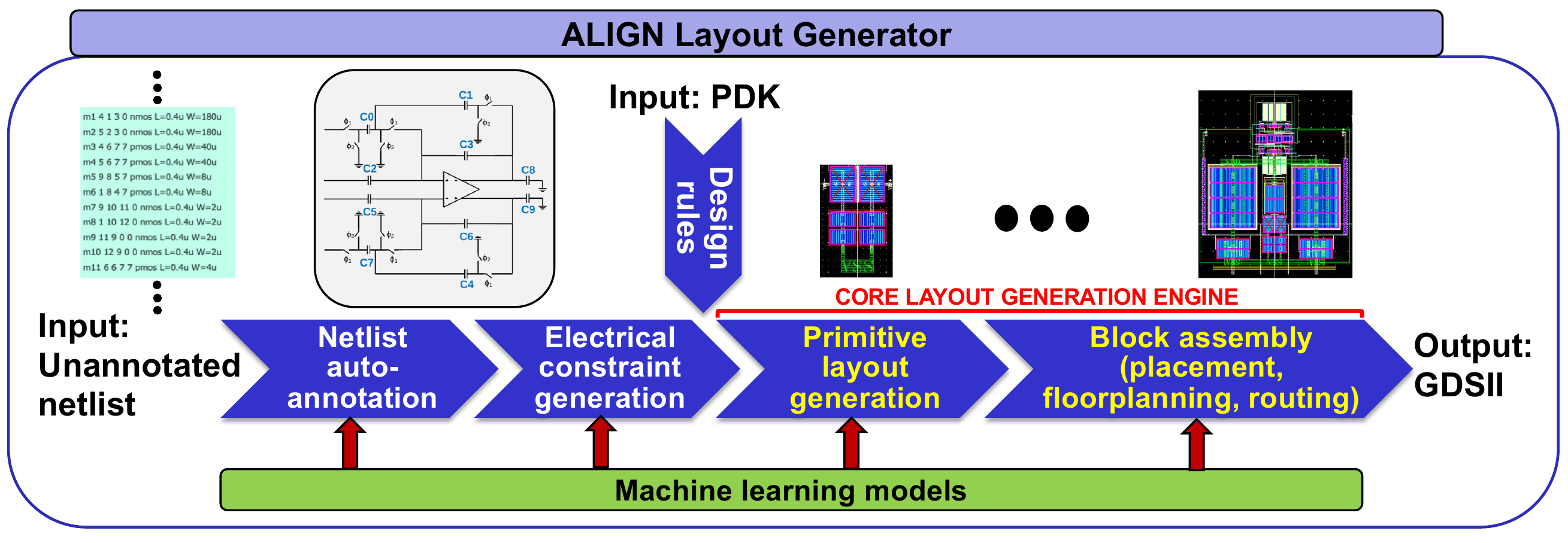}
\caption{Overview of the ALIGN flow.}
\label{fig:align_flow}
\end{figure}

\noindent
The ALIGN flow consists of five modules, illustrated in
Fig.~\ref{fig:align_flow}:\\
(1) {\bf Netlist auto-annotation} creates a multilevel hierarchical representation
of the input netlist and identifies structural symmetries in the netlist.  This is
a key step that is used to hierarchically build the layout of the circuit.\\
(2) {\bf Design rule capture} abstracts the proprietary PDK into a simplified grid,
appended with Boolean constraints as needed, that must be obeyed at all steps
during layout.\\
(3) {\bf Constraint generation} identifies the performance constraints
to be met, and transforms them into layout constraints, such as maximum
allowable net lengths, or constraints such as matching/common-centroid based on
structural information identified during auto-annotation.\\
(4) {\bf Parameterized primitive cell generation} automatically builds layouts for
primitives, the lowest-level blocks in the ALIGN hierarchy.  Primitives
typically contain a small number of transistor structures (each of which may be
implemented using multiple fins and/or fingers). A parameterized instance of a
primitive in the netlist is automatically translated to a GDSII layout in this step.\\
(5) {\bf Hierarchical block assembly} performs placement and routing on the
hierarchical circuit structure while meeting geometric and electrical
constraints.\\
The flow creates a separation between open-source code and proprietary data.
Proprietary PDK models must be translated into an abstraction that is used by
the layout generators.  Parts of the flow are driven by ML models: the
flow provides infrastructure for training these models on proprietary data.

The overall ALIGN flow is intended to support no-human-in-the-loop design.
However, the flow is modular and supports multiple entry points: for example,
the auto-annotation module could be replaced by designer annotation, and the
rest of the flow could be executed using this annotation.  The flow is flexible
to user input: for example, the user can specify new primitives, and they will
be used by the annotation module as well as the layout generator within the
flow.

\subsection{Netlist Auto-Annotation}

\noindent
This step groups transistors and passives in the input netlist into a
hierarchical set of building blocks and identifies constraints on the layout of
each block.  The input to ALIGN is a SPICE netlist that is converted to
a graph representation. Next, features of the graph are recognized, and a
circuit hierarchy is created.  If the input netlist is partitioned into
subcircuits, such information is used during recognition, but ALIGN does
not count on netlist hierarchy. Instead, hierarchies are automatically
identified and annotated.  It is important to note that the best layout hierarchy may
sometimes differ from a logical netlist hierarchy; hence, ALIGN may flatten netlist
hierarchies to build high-quality layouts.  

Analog designers typically choose from a large number for variants of each design
block, e.g., between textbooks and research papers, there are well over 100
widely used OTA topologies of various types (e.g., telescopic, folded cascode,
Miller-compensated). \blueHL{Prior methods are library-based (i.e., they match a
circuit to prespecified templates)~\cite{Eick11} or knowledge-based (i.e., they
determine block functionality using a set of encoded rules)~\cite{Harjani89},
or both~\cite{Wu15b}.}  Library-based methods require a large library, while
rule-based methods must be supported by an exhaustive knowledge base, both of
which are hard to build and maintain.  ALIGN uses two approaches for annotating
circuits blocks, both based on representing the circuit connectivity using a
graph representation:

\noindent
{\bf Machine learning based methods}:
For commonly encountered blocks, the problem of identifying blocks maps on
to whether a subgraph of the larger circuit is isomorphic to a known cell.
However, to allow for design variants, ALIGN uses {\em approximate} graph
isomorphism, enabled by the use of graph convolutional neural networks (GCNs)
that classify nodes within the circuit graph into classes (e.g., OTA nodes, LNA
nodes, Mixer nodes).  With some minimal postprocessing, it is demonstrated that
this approach results in excellent block recognition.  \blueHL{Details of the approach
are provided in~\cite{Kunal20}.}  A training set for the GCN, consisting of 1390
OTA circuits, including bias networks, is available on the ALIGN GitHub
repository.

\noindent
{\bf Graph traversal based methods}:
It is unrealistic to build a training set that covers every possible analog
block, and for blocks that lie outside the scope of the GCN training set,
we use graph-based approaches to recognize repeated structures within a circuit.
Such structures typically require layout constraints: for example,
analog-to-digital converters may use a set of binary weighted capacitors or a
set of resistors in an R-2R ladder, and these require careful placement in 
common-centroid fashion and symmetric routing.  ALIGN employs methods based
on graph traversal and approximate subgraph isomorphism to recognize these
array structures.

Once these structures are recognized in a very large circuit graph, they form a
level of hierarchy. Within these blocks, lower hierarchical levels can be
detected using conventional subgraph isomorphism methods: sub-blocks at these
levels have fewer variants and can be efficiently recognized using
library-based approaches.

\begin{figure}
\centering
\includegraphics[width=0.48\textwidth]{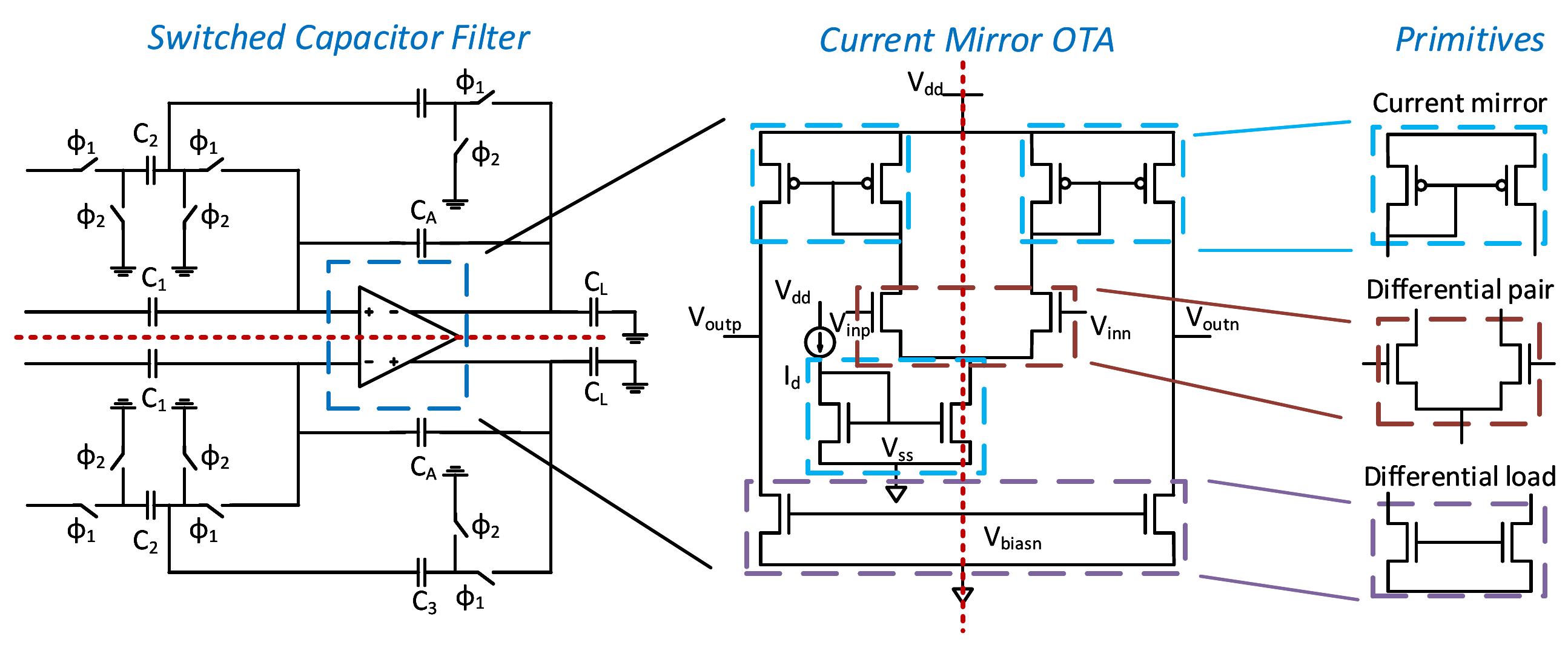}
\caption{Extracting netlist hierarchy during auto-annotation.}
\label{fig:OTA_hierarchy}
\end{figure}

Fig.~\ref{fig:OTA_hierarchy} shows the results of auto-annotation on a
switched-capacitor filter. A GCN-based approach can be used to identify the
current-mirror OTA, and then primitives within the OTA can be identified.  In the
process, lines of symmetry within each structure can be found, as illustrated
in the figure.  At the primitive level, since the layouts are generated by the
parameterized cell generator, these lines of symmetry are implicit in the
definition of the primitive. At higher levels, these can be inferred during
auto-annotation.

\subsection{Design Rule Abstraction}

\begin{figure}[th] \centering
\includegraphics[width=0.5\textwidth]{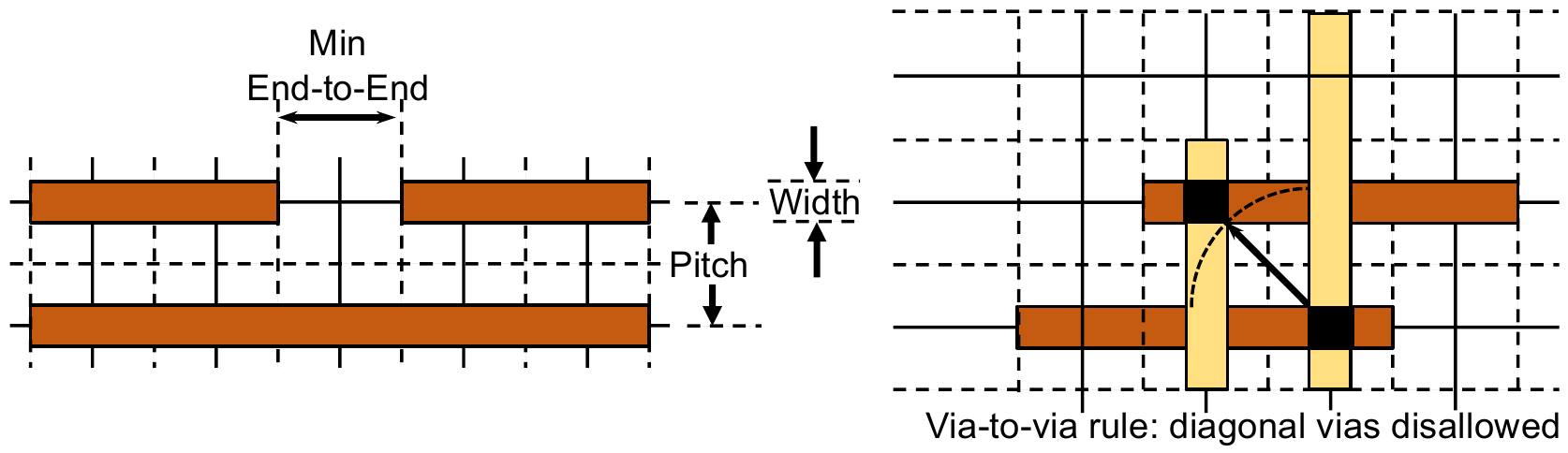} 
\includegraphics[width=0.25\textwidth]{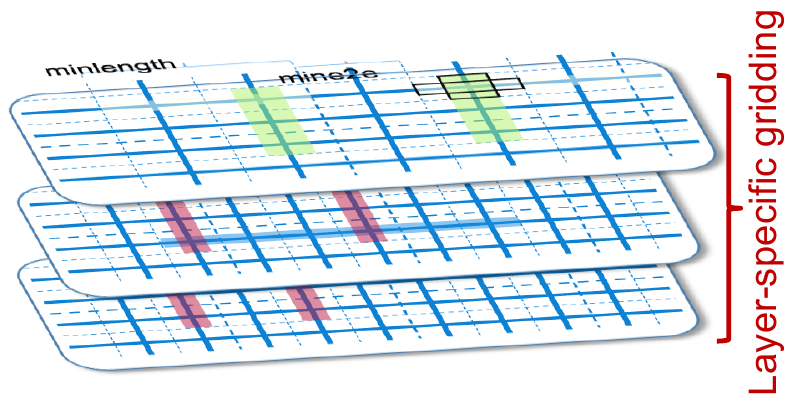} 
\caption{Design rule abstraction using per-layer grids and rules.}
\label{fig:PDKgrid}
\end{figure}

\noindent
The ALIGN layout tools are guided by process-specific design rules that ensure
DRC-correctness.  The complexity of design rules has grown significantly in
recent process generations.  Efforts at building generalized abstractions for
process rules have previously been proposed (e.g.,~\cite{Opal17}).  ALIGN uses
a more efficient design rule abstraction mechanism that creates fixed grid
structures in FEOL and BEOL layers, as illustrated in Fig.~\ref{fig:PDKgrid}.
Major grids (bold lines), represent centerlines for routes, while minor grids
(dashed lines) correspond to stopping points for features.  The gridding
structure and basic process information is abstracted into a JSON file.  For
BEOL layers, this includes:
\begin{itemize}
\item default wire dimensions, pitch, and grid offset ({\tt Pitch}, {\tt
Width}, {\tt MinL}, {\tt MaxL}, {\tt Offset}).
\item end-to-end spacing design rules ({\tt EndToEnd}).
\item metal direction, colors ({\tt Direction}, {\tt Color}).
\item via rules ({\tt Space\{X/Y\}}, {\tt Width\{X/Y\}}, {\tt VencA\_\{L/H\}},
{\tt VencP\_\{L,H\}}).
\end{itemize}

While this is superficially similar to traditional $\lambda$-rules, our
abstraction permits a different gridding structure that can vary from layer to
layer, and the use of major/minor grid lines \blueHL{that represent wire
pitches, wire overhangs, as well as the ability to incorporate via rules
through Boolean constraints. Our approach reduces the complex set of
conditions embedded in thousands of rules in a design rule manual to a
massively simplified and much smaller set, enforcing some limitations through
the choice of grids.  It is found, through comparisons with manual design, that
this leads to minimal or zero degradation in layout quality.}  Advanced
commercial process nodes (22nm, 10nm, 7nm, beyond) have been abstracted into
this simplified form.  The abstraction enables layout tools to comprehend PDK
features such as regular and irregular width and spacing grids (for each
layer), minimum end to end spacing design rules (between metals in the same
track), minimum length design rules, and enforced stopping point grids. For
convenience, the JSON file also encodes per unit parasitics for metal layers
and vias.

To facilitate further layout research, we have released design rules for Mock
PDKs based on public-domain information to abstract layout rules at a 14nm
FinFET node~\cite{Lin14} and a 65nm bulk node~\cite{Steegen05}.  While they do
not represent a real technology, they are realistic.  Validation of the design
tools on these PDKs, which can be freely shared, helps the software development
process. 

\subsection{Constraint Generation} 

Two types of constraints are generated to guide layout:

\noindent
{\bf Geometric constraints}:
As the auto-annotation step recognizes known blocks or array structures, it
associates geometric requirements with these blocks, such as symmetry,
matching, and common-centroid constraints.  For instance,
Fig.~\ref{fig:OTA_hierarchy} shows lines of symmetry in an OTA structure that
must be respected during layout.  These constraints are extracted naturally as
part of auto-annotation.  \blueHL{In contrast with prior methods that are based on
simulation-intensive sensitivity analysis~\cite{Malavasi96} or graph traversal
based exact matching to templates~\cite{Eick11}, the approach in ALIGN
method~\cite{Kunal20ICCAD} combines graph traversal methods with machine
learning based methods and is computationally efficient, capable of finding
hierarchically nested symmetry constraints even under approximate matches.}

\noindent
{\bf Electrical constraints}:
ALIGN generates a layout based on a fixed netlist, and performance shifts are
driven by changes in parasitics from netlist-level estimates to post-layout
values.  Therefore, ALIGN translates electrical constraints to bound the
maximum parasitics at any node of the circuit.  For instance, an electrical
constraint may be translated to a maximum limit on the resistance of a wire
connecting two nodes, which in turn corresponds to a constraint on the maximum
length, the number of parallel metal tracks, and the number of vias on the
route connecting these nodes.  \blueHL{This feature is currently being
implemented in ALIGN~\blueHL{\cite{Li20,Li20ICCAD}} and is a work in progress.
The essential idea is to develop a fast ML inference engine that operates
within the inner loop of an iterative placer, and for each placer
configuration, determines whether or not its electrical constraints are
satisfied.}

These constraints are passed on to the layout generation engine to guide layout
at all levels of hierarchy.

\subsection{Parameterized Primitive Layout Generation} 

\begin{figure} 
\centering
\includegraphics[width=0.4\textwidth]{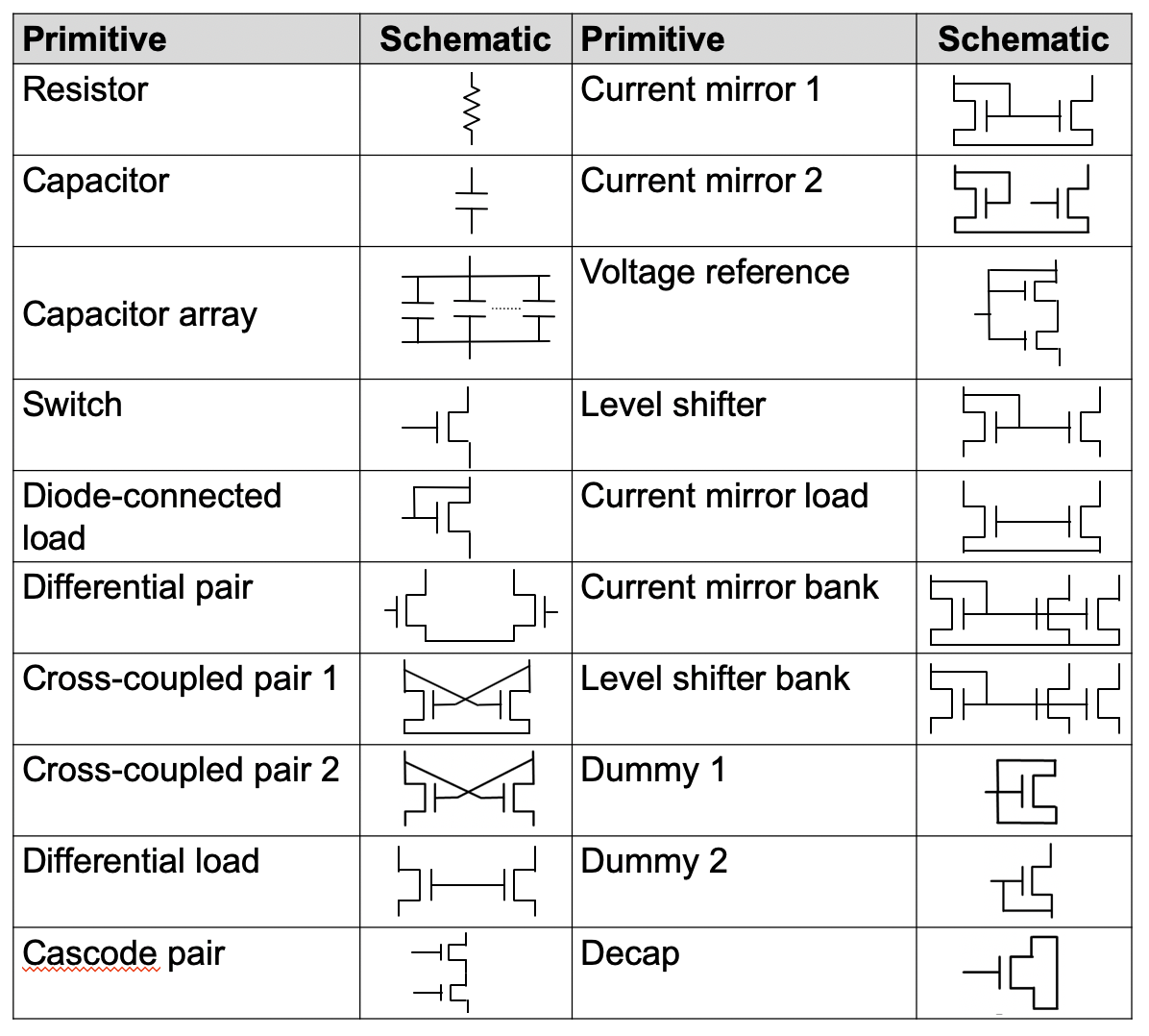}
\caption{Examples of primitive structures.}
\label{fig:PrimitiveList}
\end{figure}

ALIGN provides the user with a predefined library of parameterizable primitives,
as illustrated in Fig.~\ref{fig:PrimitiveList}.  Each primitive consists
of a small number of transistor or passive units; however, each such unit may 
consist of multiple replicated structures, such as multifin/multifinger transistors,
or resistive/capacitive arrays.  

\begin{figure}
\centering
\includegraphics[width=0.50\textwidth]{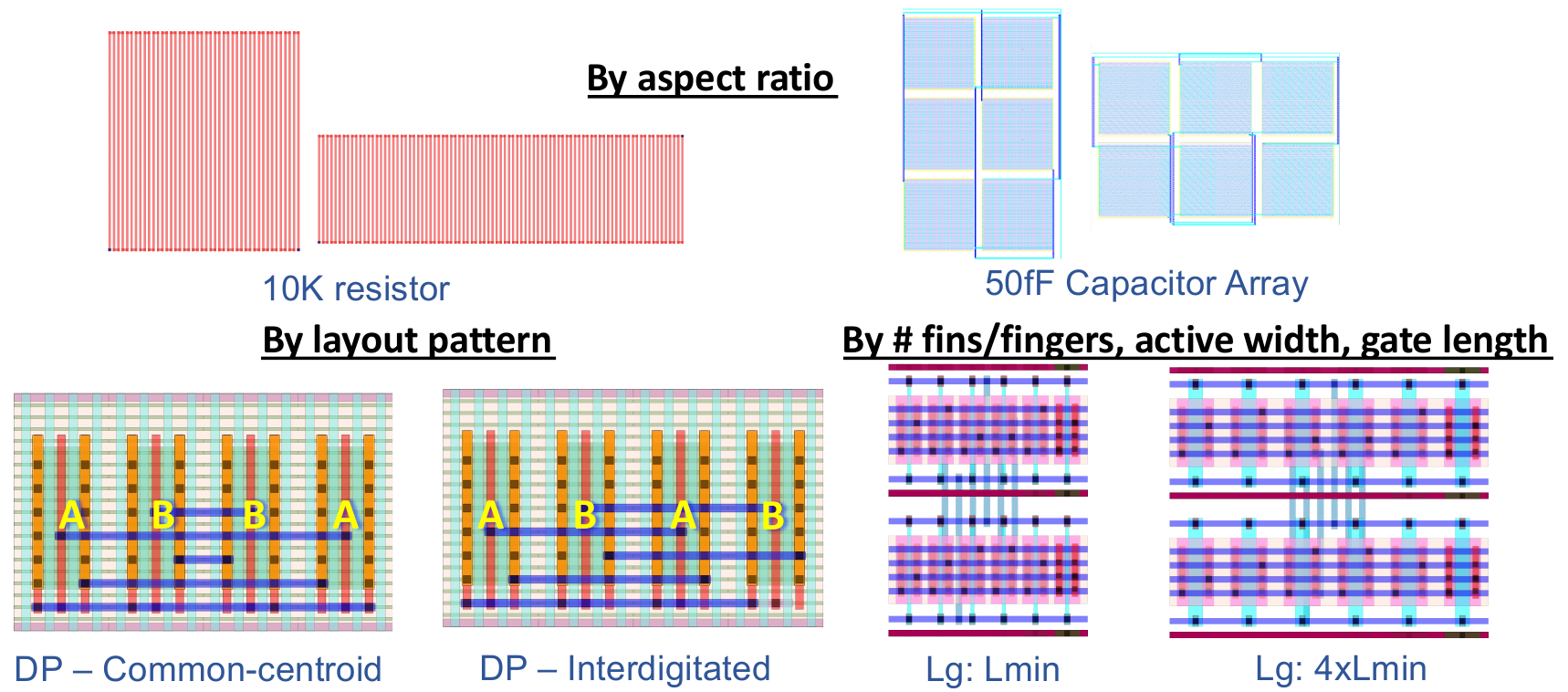}
\caption{Parameterization of primitive layouts.}
\label{fig:PrimitiveParameterization}
\end{figure}

The primitive cell layout follows the gridded abstraction defined by the design
rules, and cell generation can be parameterized in terms of the unit cell and the
number of unit cells, as shown in Fig.~\ref{fig:PrimitiveParameterization}.
For a transistor, a unit cell may be parameterized by the number of fins in a
FinFET technology; for a capacitor, parameterization may correspond to the size
of the unit capacitor. Additionally, primitive layouts can be parameterized by
their aspect ratio, their layout style (common-centroid vs. interdigitated
transistors), the gate length, the effective widths of critical wires in the
cell, etc.

The utility in recognizing primitives and creating parameterized layouts is
in enabling ALIGN to create layouts that incorporate the appropriate geometric
constraints (e.g., symmetry or common-centroid).  In principle, a layout could
be built using a ``sea of transistors,'' where the primitive corresponds to
a single transistor, but it would be challenging for such an approach to
enforce symmetry requirements beyond the transistor primitives.
\blueHL{Prior methods for primitive layout
generation~\cite{Bhattacharya04,Lourenco06,Zhang06,Yilmaz08 } have generally
not been as modular or scalable as the ALIGN approach.}

\subsection{Hierachical Block Assembly}

Given the layouts of all primitives and the hierarchical block level structure
of the circuit, extracted during auto-annotation, the placement and routing
step performs hierarchical block assembly that obeys the geometric and
electrical constraints described earlier.

Each layout block in the hierarchy can have multiple layout options with
different shapes generated for each module.  For example, primitives can be
parameterized by aspect ratio, and multiple aspect ratios for other blocks may
be generated.  Flexible shapes drive floorplanning-like placement algorithms
that deliver compact layouts under the electrical and geometric constraints
passed on to them by the constraint generation step.  Routing is integrated
into each hierarchical level, accounting for net length/parasitic constraints,
shielding and symmetry requirements, and conforming with the design rules
embedded into the PDK abstraction.  The placer is based on prior work using the
sequence pair method~\cite{Ma11} and can handle general geometric constraints,
such as symmetry, matching and alignment.  Symmetry, shielding and
resistance-constrained routing are supported during routing.

The ALIGN flow can employ one of two detailed routers: (a) A constructive
router that uses an integer linear programming formulation and an A* algorithm;
this works particularly well for more sparse designs. (b) A SAT-based detailed
router\footnote{\url{github.com/ALIGN-analoglayout/AnalogDetailedRouter}},
released by Intel, which is well suited for congested designs.

%% file: opensource.tex
\section{Working in an Open-source Environment}

\subsection{Why Open-source Software?}

\blueHL{Aside from technical innovations, ALIGN breaks new ground in providing
a fully open-source analog layout software, something that has not been
available in the past.  The availability of open-source software is crucial for
nurturing future innovations in the field.  First, further research can build
upon a ``piece of the puzzle'' of analog layout design: for instance, a new
cell generator can plug into the open-source ALIGN flow and show end-to-end
results from netlist to layout, rather than providing limited results at the
end of cell generation.  Second, open-source enables a path to ensure that
reported results can be reproducible.  The traction for open-source is
evidenced not only through the efforts in ALIGN, but also in other notable
efforts on analog layout~\cite{Xu19ICCAD}, digital layout (including back-end
infrastructure such as parasitic extraction on power delivery that is more
broadly applicable to any other class of design~\cite{Ajayi19}.}

\subsection{Open-source Designs}
\label{sec:opensource}

\noindent
Unlike digital designs, where a wealth of designs exists in the public domain,
the font of analog designs is very sparse. Design parameters tend to be closely
linked with process nodes and existing automation flows do not allow robust
circuit optimization to meet constraints.  Sharing designs based on a
commercial PDK over multiple institutions requires a multiway nondisclosure
agreement involving the institutions, the foundry, and the foundry access
provider. Within the ALIGN team, this issue was complicated by the need for
such an agreement to cover both academic and industry team members.

The ALIGN GitHub repository hosts a number of sized analog netlists, a set that
is growing, to facilitate open research.  These netlists contain testbenches
that measure the performance parameters of the circuit to verify its adherence
to specifications. Moreover, as stated earlier, the repository contains unsized
netlist topologies for a variety of OTA circuits.

%% file: software.tex
\subsection{Software Infrastructure}

The software flow is maintained on a GitHub repository~\cite{ALIGN-public} and
may be downloaded and installed in a native Linux environment.  Alternatively, 
it may be run in a lightweight Docker container that performs operating system
virtualization and enables portability and ease of maintenance.  ALIGN can
leverage the use of other open-source tools such as the KLayout layout viewer.
The core software flow is Python-based, and the computationally intensive
engines -- notably the placer and router -- are implemented in C++.

The project is aided by the use of tools that are vital to a open-source
infrastructure with continuous integration (CI). These include CI build flows,
using CircleCI, for automated build of new components as they are added to the
repository; unit testing, using pytest, to verify the correctness of individual
units of source code that is added to the repository; code coverage to measure
how much of the code is executed by the automated tests, using coverage.py with
Codecov for tracking; and automated code review for code quality checks using
Codacy.

%% file: results.tex
\begin{figure}
\centering
\subcaptionbox{SC DC2DC\label{fig:SCDC2DC}}[1in]{%
	 \includegraphics[height=1.5in]{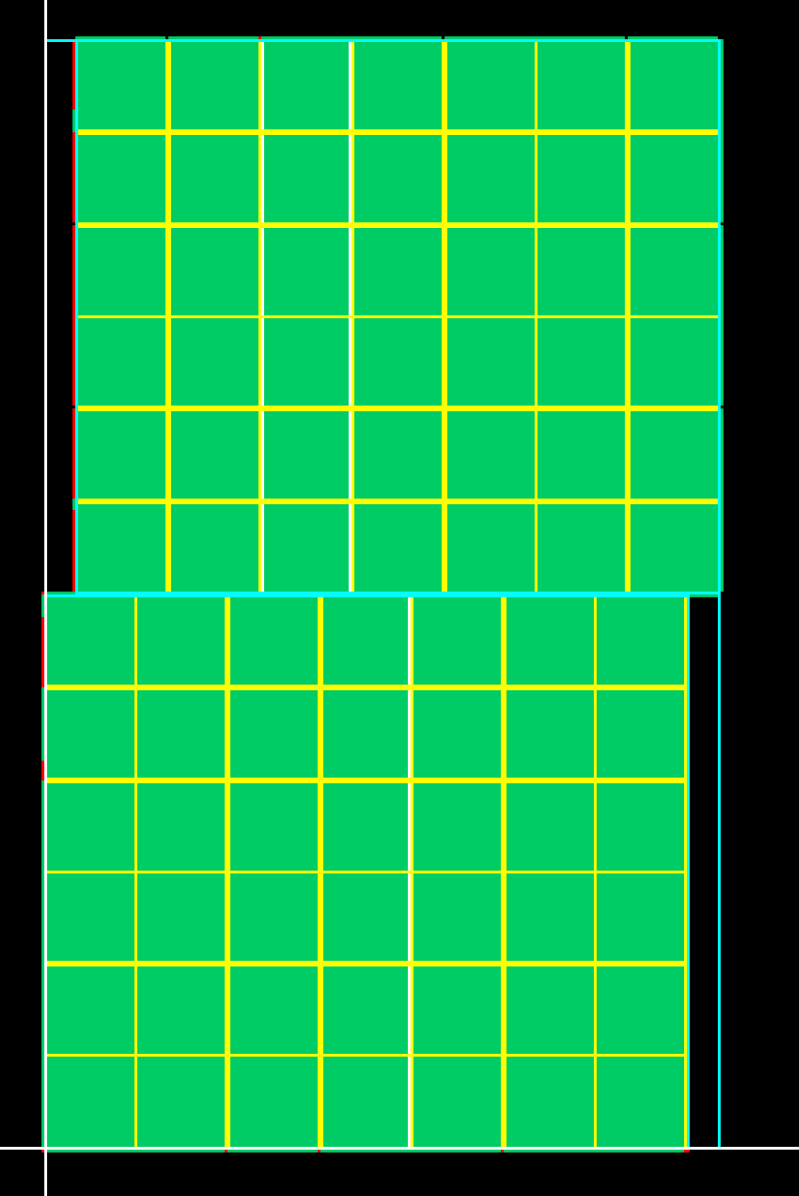}
}
\subcaptionbox{OTA with bias circuitry\label{fig:OTA}}{%
	 \includegraphics[height=1.5in]{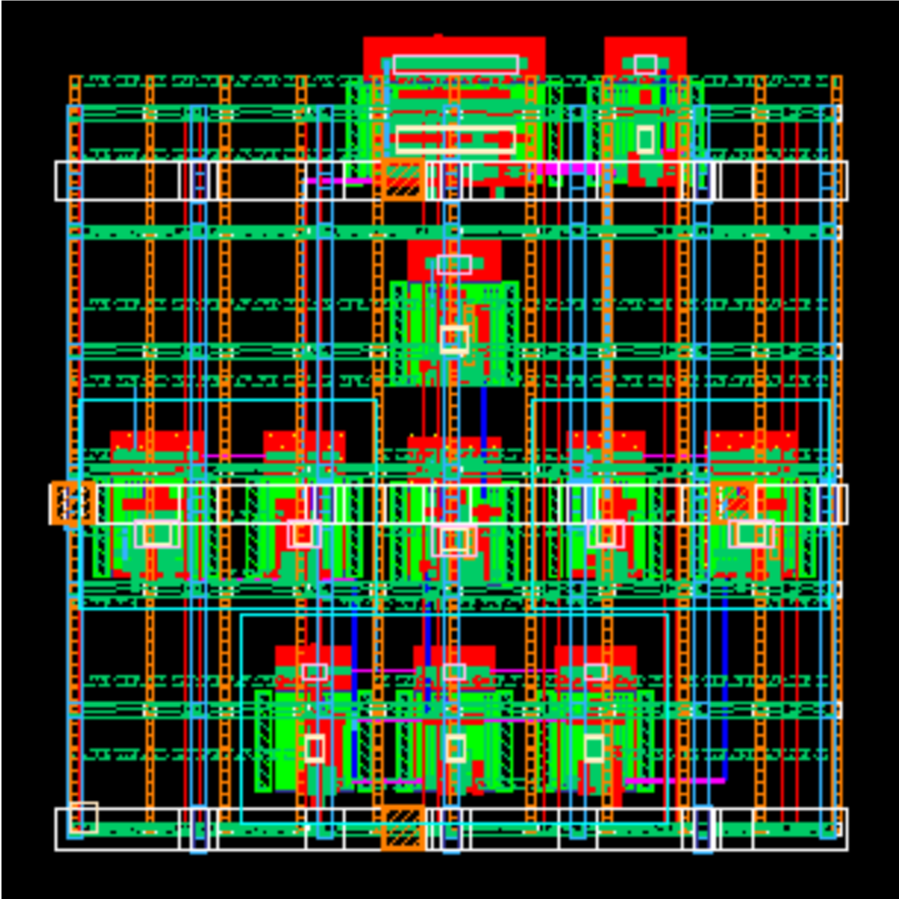}
}

\subcaptionbox{SC Filter\label{fig:SCFilter}}{%
	 \includegraphics[height=1.28in]{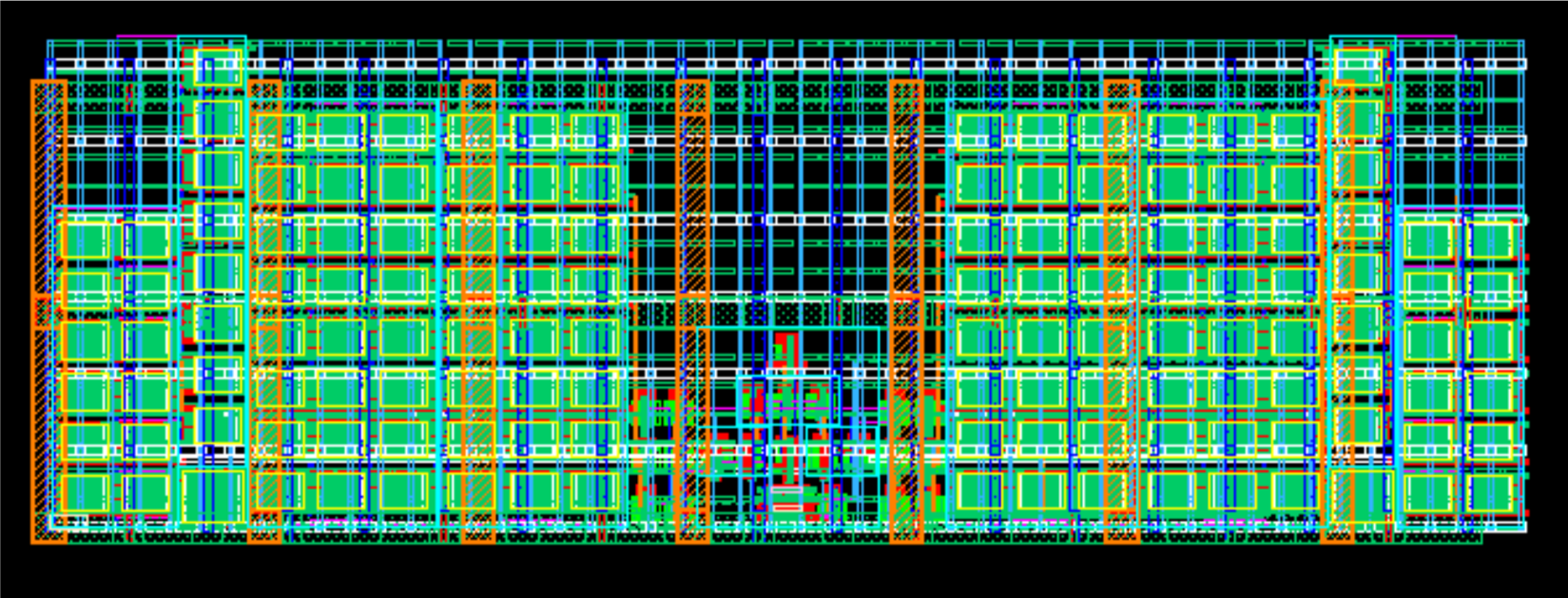}
}

\subcaptionbox{ADC\label{fig:UWADC}}{%
	\includegraphics[height=1.8in]{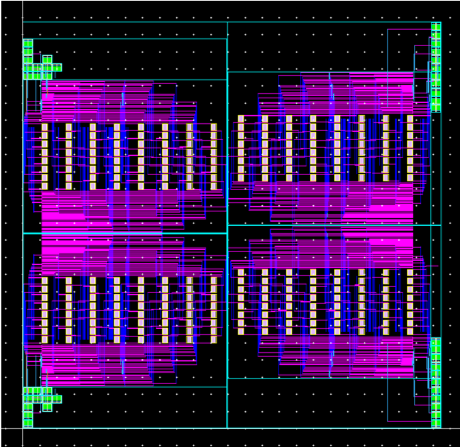}
}
\subcaptionbox{BPF\label{fig:BPF}}{%
	\includegraphics[height=1.8in]{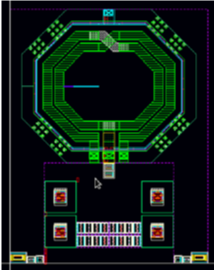}
}
\subcaptionbox{Equalizer\label{fig:Equalizer}}{%
	 \includegraphics[height=1.5in]{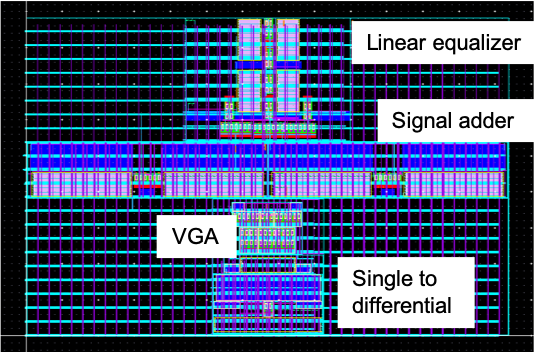}
}
\subcaptionbox{Optical Receiver\label{fig:OpticalReceiver}}{%
	 \includegraphics[height=1.7in]{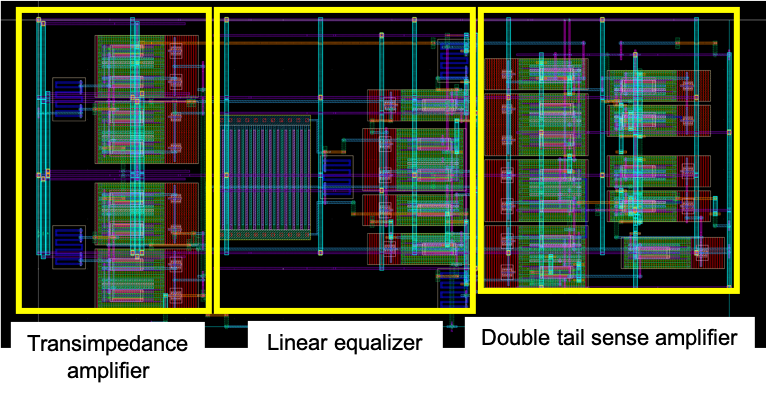}
}
\caption{Sample layouts generated by ALIGN. Note that the block sizes are
different; the layouts are not on the same scale.}
\label{fig:Layouts}
\end{figure}

\ignore{
\begin{table}[t]
\centering
\caption{Post-Layout Performance Analysis of the ALIGN-generated OTA \redHL{\bf{OLD}}}
\label{tbl:OTA_performance}
\hspace*{-0.1in}
\begin{tabular}{|l|r|r|r|}
\hline
                    & Schematic & Post-Layout     & Post-Layout      \\
                    &           & (C extract)     & (RC extract)     \\
\hline \hline
Gain (dB)           & 39.30     & 39.29  ($0$\%)  & 37.25  ($-5$\%)  \\
\hline
3dB frequency (MHz) & 10.65     & 10.47  ($-2$\%) & 10.47  ($-2$\%)  \\
\hline
UGF (MHz)           & 493.62    & 489.85 ($-1$\%) & 382.95 ($-22$\%) \\
\hline
Input offset (mV)   & 0         & 0               & 0.15             \\
\hline
\end{tabular}
\end{table}

\begin{table}[t]
\centering
\caption{Post-Layout Performance Analysis of the ALIGN-generated
Switched-Capacitor Filter Layout \redHL{\bf{OLD}}}
\label{tbl:SCF_performance}
\begin{tabular}{|l|r|r|r|}
\hline
Specification     & Schematic & Post-Layout \\ 
                  &           & (C extract) \\ \hline \hline
Inband gain (dB)  & 5.53      & 5.48 (-1\%) \\ \hline
Input Offset (mV) & 0         & 0           \\ \hline
\end{tabular}
\end{table}
}

\begin{table}[htb]
\centering
\caption{\blueHL{Post-Layout Performance Analysis of the ALIGN-generated OTA}}
\label{tbl:OTA_performance}
\hspace*{-0.1in}
\resizebox{\linewidth}{!}{
\begin{tabular}{|l|c|c|c|}
\hline
                        & Schematic  & Manual layout  & ALIGN Layout \\
                        &            & (RC extract)   & (RC extract) \\
\hline \hline
Gain (dB)               & 24.28      & 24.22          & 24.14        \\
\hline
3dB frequency (MHz)     & 24         & 24             & 24           \\
\hline
UGF (MHz)               & 199        & 197            & 198          \\
\hline
Phase margin ($^\circ$) & 89         & 88             & 88           \\
\hline
Input offset (mV)       & $\sim$0    & 0.13           & 0.10         \\
\hline
\end{tabular}
}
\end{table}

\begin{table}[htb]
\centering
\caption{\blueHL{Post-Layout Performance Analysis of the ALIGN-generated
Switched-Capacitor Filter Layout}}
\label{tbl:SCF_performance}
\resizebox{\linewidth}{!}{
\begin{tabular}{|l|r|r|r|r|}
\hline
Specification              & Schematic & Manual layout & ALIGN layout \\ 
                           &           & (RC extract)  & (RC extract) \\ \hline \hline
Gain (dB)                  & 16.1      & 15.84         & 15.59        \\ \hline
3dB frequency (KHz)        & 503       & 511           & 524          \\ \hline
Unity gain frequency (KHz) & 3435      & 3415          & 3610         \\ \hline
Input offset (mV)          & 0         & 0.13          & 0.10         \\ \hline
\end{tabular}
}
\end{table}

\begin{table}[b]
\caption{Comparing the performance of the schematic (S), manual layout (M), and 
the ALIGN-generated layout (A)}
\label{tbl:manualcomparison}
\resizebox{\linewidth}{!}{
\begin{tabular}{|l||c|c||c|c|}
\hline
  & \multicolumn{2}{c||}{SDC} & \multicolumn{2}{c|}{Signal adder}  \\ \cline{2-5}
  & Gain (dB) [$\Delta$Gain]          & BW (GHz) [$\Delta$BW]
  & Gain (dB) [$\Delta$Gain]          & BW (GHz) [$\Delta$BW]      \\ \hline
S & -$5.6$                            & $27.9$ 
  & $2.9$                             & $24.5$                     \\ \hline
M & -$6.0$ [-$3.5\%$]                 & $19.8$ [-$29.0\%$]
  & $2.3$ [-$5.8$\%]                  & $15.7$ [-$36.1$\%]         \\ \hline
A & -$6.1$ [-$4.8$\%]                 & $23.0$ [-$17.6\%$]
  & $2.2$ [-$7.0$\%]                  & $19.8$ [-$19.2$\%]         \\ \hline \hline
  & \multicolumn{2}{c||}{VGA}         & \multicolumn{2}{c|}{Linear equalizer} \\ \cline{2-5}
  & Gain (dB) [$\Delta$Gain]          & BW (GHz) [$\Delta$BW]
  & Gain (dB) [$\Delta$Gain]          & BW (GHz) [$\Delta$BW]      \\ \hline
S & -$10.9$dB $\sim$ $5.5$            & $26.25$
  & $1.2$                             & $18.0$                     \\ \hline
M & -$10.9$dB $\sim$ $5.6$ [$1.0$\%]  & $12.58$ [-$52.1$\%]
  & $0.9$ [-$3.4$\%]                  & $13.9$ [-$22.8$\%]         \\ \hline
A & -$11.0$dB $\sim$ $5.0$ [-$5.6$\%] & $13.40$ [-$49.0$\%]
  & $0.8$ [-$4.2$\%]                  & $15.7$ [-$12.8$\%]         \\ \hline
\end{tabular}
}
\end{table}

\section{Results}
\label{sec:results}

The ALIGN flow has been applied to generate layouts for circuits that
lie in all four classes: low-frequency analog, wireline, wireless, and power
delivery.  \blueHL{We are unaware of a prior layout generator that has been
demonstrated to handle such a broad class of circuits.}  Fig.~\ref{fig:Layouts}
illustrates a sample set of layouts generated using ALIGN: these include a
current-mirror OTA with bias circuitry and its power grid (Fig.~\ref{fig:OTA}), a switched
capacitor (SC) filter containing the OTA~(Fig.~\ref{fig:SCFilter}), an
analog-to-digital converter [all low-frequency analog], a bandpass
filter~(Fig.~\ref{fig:BPF}) [wireless], a switched capacitor DC-to-DC
converter~(Fig.~\ref{fig:SCDC2DC}) [power delivery], and an
equalizer~(Fig.~\ref{fig:Equalizer}) and an optical
receiver~(Fig.~\ref{fig:OpticalReceiver}) [both wireline].  The layouts are
compact and regular.

A set of representative results for the post-layout performance analysis of
ALIGN-generated layouts for the OTA (Fig.~\ref{fig:OTA}) and the
switched-capacitor filter (Fig.~\ref{fig:SCFilter}) containing the OTA are
shown in Tables~\ref{tbl:OTA_performance} and~\ref{tbl:SCF_performance},
respectively.  For the larger block, the switched-capacitor filter, the
extraction results show a good match with the schematic simulation (this level
of mismatch between schematic and layout performance is quite normal in analog
design), attesting to the quality of the layout.  Moreover, the layout respects
symmetry constraints that are considered important by analog designers to guard
against parasitic mismatch due to systematic variability.  \blueHL{For both
layouts, the performance of the ALIGN-generated layout is very close to that of
the manual layout.}

For a set of wireline circuits, Table~\ref{tbl:manualcomparison} shows a
comparison between the performance of the ALIGN-generated layout and a
hand-crafted manual layout, and shows that the performance of
both layouts is comparable.

%% file: conclusion.tex
\section{Conclusion}

This paper summarizes the current state of the ALIGN flow for automated analog
layout synthesis.  ALIGN is open-source and may be downloaded and used
freely~\cite{ALIGN-public}.  Currently, the project has seen about 24 months of
development, and can already synthesize layouts for a wide variety of analog
circuits. It is expected that the capabilities of ALIGN will be enhanced
significantly over the next few years, handling more sophisticated circuits,
more complex constraints, and improved software robustness.  The inherent
hierarchical approach adopted by ALIGN is key to ensuring scalability of the
software to larger designs in future, while also providing high-quality
solutions.